\documentclass[aps, prl, twocolumn, groupedaddress, 10pt, showpacs ,superscriptaddress]{revtex4-1}

\usepackage{graphicx}
\usepackage{color}
\usepackage{amsfonts, amsmath, amsthm,amssymb, amscd}

\newcommand{\ftk}[2]{\text{FT}\big[#1\big](#2)}
\newcommand{\ft}[1]{\text{FT}\big[#1\big]}
\newcommand{\iftx}[2]{\text{FT}^{-1}\big[#1\big](#2)}

\def\epp{\: .}
\def\epc{\: ,}
\def\lam{\lambda}
\def\blam{{\boldsymbol{\lambda}}}
\def\brho{{\boldsymbol{\rho}}}
\def\limth{\lim\nolimits_\text{th}}
\def\cO{\mathcal{O}}

\begin{document}

\title{Quenching the Anisotropic Heisenberg Chain: Exact Solution\\ and Generalized Gibbs Ensemble Predictions}

\author{B. Wouters}
\affiliation{Institute for Theoretical Physics, University of Amsterdam, Science Park 904\\
Postbus 94485, 1090 GL Amsterdam, The Netherlands}

\author{J. De Nardis}
\affiliation{Institute for Theoretical Physics, University of Amsterdam, Science Park 904\\
Postbus 94485, 1090 GL Amsterdam, The Netherlands}

\author{M. Brockmann}
\affiliation{Institute for Theoretical Physics, University of Amsterdam, Science Park 904\\
Postbus 94485, 1090 GL Amsterdam, The Netherlands}

\author{D. Fioretto}
\affiliation{Institute for Theoretical Physics, University of Amsterdam, Science Park 904\\
Postbus 94485, 1090 GL Amsterdam, The Netherlands}

\author{M. Rigol}
\affiliation{Department of Physics, The Pennsylvania State University,
University Park, Pennsylvania 16802, USA}

\author{J.-S. Caux}
\affiliation{Institute for Theoretical Physics, University of Amsterdam, Science Park 904\\
Postbus 94485, 1090 GL Amsterdam, The Netherlands}

\date{\today}

\begin{abstract}
We study quenches in integrable spin-1/2 chains in which we evolve the ground state of the antiferromagnetic Ising 
model with the anisotropic Heisenberg Hamiltonian. For this nontrivially interacting situation, an application of the 
first-principles-based quench action method allows us to give an exact description of the postquench steady state 
in the thermodynamic limit. We show that a generalized Gibbs ensemble, implemented using all known 
local conserved charges, fails to reproduce the exact quench action steady state and to correctly predict 
postquench equilibrium expectation values of physical observables.
This is supported by numerical linked-cluster calculations within the diagonal 
ensemble in the thermodynamic limit. 
\end{abstract}


\maketitle

\paragraph*{Introduction.}

Out-of-equilibrium phenomena are of importance throughout physics, in fields ranging from cosmology 
\cite{1980_Kibble_PR_67} and superfluid helium \cite{1985_Zurek_NATURE_317}, heavy-ion collisions 
\cite{2005_Moore_PRC_71}, pattern formation \cite{1993_Cross_RMP_65}, exclusion processes \cite{2011_Chou_RPP_74}, and glasses \cite{2005_Kurchan_NATURE_433} all the way to atomic-scale isolated quantum systems 
\cite{2011_Polkovnikov_RMP_83}. Much recent experimental and theoretical activity has been focused on the latter, 
raising fundamental questions as to whether, how and to what state such systems relax under unitary time evolution 
following a sudden quantum quench \cite{2006_Calabrese_PRL_96, 2007_Rigol_PRL_98, 2006_Rigol_PRA_74, rigol_dunjko_08,
2008_Barthel_PRL_100, 2010_Cramer_NJP_12, 2011_Cassidy_PRL_106, 2011_Calabrese_PRL_106, 2012_Calabrese_JSTAT_P07016, 2012_Calabrese_JSTAT_P07022, 
2013_Pozsgay_JSTAT_P07003, 2014_Fagotti_PRB_89, 2014_Bucciantini_JPA_47, 2009_Barmettler_PRL_102, 2010_Barmettler_NJP_12, 
2009_Rossini_PRL_102, 2010_Rossini_PRB_82, 2009_Faribault_JMP_50, 2010_Mossel_NJP_12, 2011_Igloi_PRL_106, 
2011_Banuls_PRL_106, 2013_Liu, 2011_Rigol_PRA_84, 2012_Brandino_PRB_85, 2012_Demler_PRB_86, 2012_He_PRA_85, 
2013_He_PRA_87, 2013_Heyl_PRL_110, 2013_Pozsgay_Jstat_10, 2014_Fagotti, 2014_Heyl, 2013_Marcuzzi_PRL_111, 
2013_Mussardo_PRL_111, 2013_Kormos_PRB_88, 2014_Essler_PRB_89}. From this work, two scenarios for equilibration have emerged, 
one applicable to models having only a few local conserved quantities, the other relevant to integrable models characterized 
by an infinite number of local conserved charges. In the former, thermalization to a Gibbs ensemble is the 
rule \cite{rigol_dunjko_08}, while in the latter, equilibration to a so-called generalized Gibbs ensemble (GGE) 
\cite{2007_Rigol_PRL_98,2006_Rigol_PRA_74} is generally thought to occur, in particular for lattice spin systems 
\cite{2008_Barthel_PRL_100, 2010_Cramer_NJP_12, 2011_Cassidy_PRL_106,2011_Calabrese_PRL_106, 2012_Calabrese_JSTAT_P07016, 
2012_Calabrese_JSTAT_P07022, 2013_Pozsgay_JSTAT_P07003, 2014_Fagotti_PRB_89, 2014_Bucciantini_JPA_47}. 

In this Letter, we study a quench in which the second scenario breaks down. Our initial state, defined as a purely antiferromagnetic (spin-1/2 N{\'e}el) state, is let to evolve unitarily in time according to the XXZ spin chain Hamiltonian. This is a physically meaningful quench protocol, which can, in principle be implemented using cold atoms \cite{2006_Kinoshita_NATURE_440, 2012_Trotzky_NATPHYS_8,2012_Cheneau_NATURE_481,2012_Gring_SCIENCE_337, 2013_Fukuhara_NATURE_502}. We provide a thermodynamically exact solution for the steady state reached long after the quench, derived directly from microscopics using the recently proposed quench action method \cite{2013_Caux_PRL_110}. The solution takes the form of a set of distributions of quasimomenta that completely characterizes the macrostate representing the steady state, from which observables of interest can be calculated. As a stringent test, it correctly reproduces the expectation values of all local conserved charges. Furthermore, we implement a numerical linked-cluster expansion (NLCE) \cite{rigol_14,rigolpre_14} whose results support the correctness of the quench action approach. Our application of the latter to nontrivially interacting lattice models follows up on the recent quench action solution of interaction quenches in one-dimensional Bose systems \cite{2014_DeNardis_PRA_89} and demonstrates the broad applicability of the approach.

Besides providing the exact solution using the quench action, we explicitly construct
a GGE for the N\'eel-to-XXZ quench using all known local conserved charges, enabling an analytical check of the GGE logic applied to interacting systems. We show that it fails to reproduce the steady state as predicted by the quench action. As a consequence, equilibrium expectation values of physical observables are predicted differently by the quench action method, which corresponds to the prediction of the diagonal ensemble, and the GGE based on all known local conserved charges. We display these differences explicitly for short-distance spin-spin correlations and verify them using NLCE. Our results highlight how far-from-equilibrium dynamics can reveal the effects of physically relevant but unknown conserved quantities in interacting integrable models.

\paragraph*{Quench protocol.}
Our initial state is the ground state of the antiferromagnetic Ising model, namely, the translationally invariant N{\'e}el state
\begin{equation}\label{eq:Neel_state}
	|\Psi_0\rangle = \frac{1}{\sqrt{2}}\left(\left|\uparrow\downarrow \uparrow\downarrow\ldots\right\rangle+
	\left|\downarrow\uparrow\downarrow\uparrow \ldots\right\rangle\right).
\end{equation}  
The time evolution after the quench is governed by the antiferromagnetic XXZ spin chain Hamiltonian
\begin{equation}\label{eq:Hamiltonian_XXZ}
	H = \frac{J}{4}\sum_{j=1}^{N}\left[\sigma_{j}^{x}\sigma_{j+1}^{x}+\sigma_{j}^{y}\sigma_{j+1}^{y}+
	\Delta ( \sigma_{j}^{z}\sigma_{j+1}^{z}-1)\right],
\end{equation}
with exchange coupling $J>0$. The N{\'e}el state is the ground state in the limit $\Delta \rightarrow \infty$. 
The Pauli matrices $\sigma_j^\alpha$ ($\alpha=x,y,z$) represent the spin-$1/2$ degrees of freedom at lattice 
sites $j=1,2,\ldots, N$, and we assume periodic boundary conditions $\sigma_{N+1}^\alpha = \sigma_1^\alpha$.
We restrict our analysis to quenches for which $\Delta \geq 1$ (details for the $\Delta = 1$ case are provided in Ref.~\cite{FollowUp}).

Eigenstates of the Hamiltonian \eqref{eq:Hamiltonian_XXZ} can be obtained by Bethe ansatz \cite{1931_Bethe_ZP_71,1958_Orbach_PR_112}. 
Each normalized Bethe wave function
\begin{equation}\label{eq:BA_state}
	|\blam\rangle = \sum_{\boldsymbol{x}} \sum_{Q}A_Q(\blam)\prod_{j=1}^M e^{i x_j p(\lambda_{Q_j})} 
	\sigma_{x_j}^-\left|\uparrow\uparrow\ldots\uparrow\right\rangle
\end{equation}
lies in a fixed magnetization sector $\langle\sigma_\text{tot}^z\rangle/2 = N/2-M$. 
It is completely specified by a set of complex quasimomenta or rapidities $\boldsymbol{\lam}= \{\lambda_k\}_{k=1}^M$, 
which satisfy the Bethe equations 
\begin{equation}\label{eq:BAE}
	\left(\frac{\sin(\lambda_j+i\eta/2)}{\sin(\lambda_j-i\eta/2)}\right)^N=
	-\prod_{k=1}^M\frac{\sin(\lambda_j-\lambda_k+i\eta)}{\sin(\lambda_j-\lambda_k-i\eta)},
\end{equation}
for $j=1,\ldots,M$. The parameter $\eta>0$ is related to the anisotropy parameter $\Delta=\cosh(\eta)$. 
The first sum in Eq.~\eqref{eq:BA_state} is over all ordered configurations 
$\boldsymbol{x}= \{x_j\}_{j=1}^M \subset \{1,\ldots,N\}$ of down spin positions, 
while the second sum runs over all permutations $Q$ of labels $\{ 1, ..., M\}$. 
$A_Q(\blam)$ are rapidity-dependent amplitudes \cite{1931_Bethe_ZP_71,1958_Orbach_PR_112}. The total momentum 
and energy of a Bethe state are given by 
\begin{align}
	P_\blam &=\sum_{j=1}^M p(\lam_j) \epc\quad p(\lam) = i \ln\left[\frac{\sin(\lambda - 
	i \eta/2)}{\sin(\lambda + i \eta/2)}\right]\epc \\ 
	\label{eq:total_energy}
	\omega_{\blam} &= \sum_{j=1}^M e(\lam_j)\epc\quad e(\lam) = - J \pi  \sinh(\eta) a_1(\lambda) \epc
\end{align}
where $a_1(\lambda) = \sinh(\eta)/[\pi (\cosh\eta -\cos 2\lambda)]$.

Bethe states are classified according to the string hypothesis \cite{1931_Bethe_ZP_71, 1971_Takahashi_PTP_46}. 
Rapidities arrange themselves in strings $\lam^{n,a}_\alpha=\lam^n_\alpha+\tfrac{i\eta}{2}(n+1-2a) + 
i\delta^{n,a}_\alpha$, $a=1,\ldots,n$, where $n$ is the length of the string and the deviations 
$\delta^{n,a}_\alpha$ vanish (typically exponentially) upon taking the infinite-size limit. 
For $\Delta>1$, the string centers $\lam^n_\alpha$ lie in the interval $[-\pi/2,\pi/2)$. Physically, 
such an $n$-string corresponds to a bound state of $n$ magnons, which in the Ising limit $\Delta\to\infty$ 
can be seen as a block of $n$ adjacent down spins. 

At time $t$ after the quench, the state of the system can be expanded in the basis of Bethe states
such that the postquench time-dependent expectation value of a generic operator $\cO$ is exactly given 
by the double sum
\begin{equation} \label{eq:expect1}
\left\langle \Psi (t) \right| \cO \left| \Psi (t) \right\rangle = 
\sum_{\blam,\blam'} e^{-S_{\blam}^*-S_{\blam'}}e^{i(\omega_{\blam} - 
\omega_{\blam'})t} \langle \blam| \cO |\blam'\rangle,
\end{equation}
with overlap coefficients $S_{\blam} = - \ln \left\langle \blam | \Psi_{0} \right\rangle$.

\paragraph*{Quench action.}
The double sum over the full Hilbert space in Eq.~\eqref{eq:expect1} represents a substantial bottleneck, 
its size growing exponentially with $N$. The quench action method \cite{2013_Caux_PRL_110, 2014_DeNardis_PRA_89} 
gives a handle on this double sum in the thermodynamic limit $N\to\infty$ (with $M/N=1/2$ 
fixed), denoted by $\limth$. In this limit, a state is characterized by the distributions of its string centers. 
They are given by a set of positive, smooth, and bounded densities $\boldsymbol{\rho}= \{\rho_n\}_{n=1}^\infty$ 
for the string centers $\lambda_\alpha^n$, representing a set of Bethe states with Yang-Yang (YY) entropy 
\begin{equation} \label{eq:YYentropyXXZ}
\frac{S_{\text{YY}}\! \left[ \boldsymbol{\rho}\right]}{N} = \! \sum_{n=1}^{\infty}  \int_{-\pi/2}^{\pi/2}\!\!\!\! 
d\lam \! \left[ \rho_{n} \ln(1+\eta_n) + \rho_{n,h}\ln(1+\eta_n^{-1}) \right].
\end{equation}
Here, $\rho_{n,h}$ is the density of holes of $n$-string centers \cite{KorepinBOOK, TakahashiBOOK}, 
$\eta_{n} = \rho_{n,h}/\rho_{n}$, and we leave the $\lam$ dependence implicit. The Bethe Eqs.~\eqref{eq:BAE} become a set of 
coupled integral equations \cite{1971_Takahashi_PTP_46} for the densities $\boldsymbol{\rho}$,
\begin{equation}\label{eq:BTGthlim_fact}
\rho_{n}(1 + \eta_{n}) = s \ast (\eta_{n-1}\rho_{n-1} + \eta_{n+1}\rho_{n+1}) \epc \quad n\geq 1\epc
\end{equation}
with $\eta_0(\lam)=1$ and $\rho_{0}(\lam)=\delta(\lam)$. The convolution $\ast$ is defined by
$(f\ast g)\, (\lam) = \int_{-\pi/2}^{\pi/2} f(\lam-\mu)g(\mu) d\mu $, 
and the kernel in Eqs.~\eqref{eq:BTGthlim_fact} is 
$s(\lam) = (2\pi)^{-1}\sum_{k\in\mathbb{Z}} \left[ e^{-2i k\lam}/\cosh(k\eta) \right]$.

As explained in Ref.~\citep{2014_DeNardis_PRA_89}, for a large class of physical observables, the double sum 
in Eq.~\eqref{eq:expect1} can be recast in the thermodynamic limit as a functional integral over the root 
densities $\boldsymbol{\rho}$.
The weight of the functional integral $e^{ -S_{\text{QA}}[\boldsymbol{\rho}] } $
is given by the quench action (QA) $S_{\text{QA}}[\boldsymbol{\rho}] = 
2 S[\boldsymbol{\rho}] - S_{\text{YY}}[\boldsymbol{\rho}]$, where 
$S[\boldsymbol{\rho}] = \limth \text{Re}\, S_\blam$ is the extensive real part of the 
overlap coefficient in the thermodynamic limit.
Since the quench action is extensive, real and boun\-ded from below, a saddle-point approximation 
becomes exact in the thermodynamic limit. 
At long times after the quench, 
the system relaxes to a steady state $\brho^\text{sp}$ determined by the variational equations
\begin{equation}\label{eq:variation}
0 = \left. \frac{\delta S_{\text{QA}}\left[ \brho \right] }{\delta \rho_n (\lam) } \right|_{\brho=\brho^\text{sp}}  
\qquad \text{for} \quad n\geq 1 \epp
\end{equation}
Steady-state expectation values of physical observables can then be effectively computed on this state:
\begin{equation}
\lim_{t \to \infty}\limth \left\langle \Psi (t) \right| \cO \left| \Psi (t) \right\rangle  = 
\left\langle\brho^\text{sp} \right| \cO \left| \brho^\text{sp}\right\rangle \epp
\label{eq:spev}
\end{equation}
The saddle-point distributions of string centers $\brho^\text{sp}$ thus encode all equilibrium expectation 
values and correlators of physical observables after the quench \cite{2011_Cassidy_PRL_106,2013_Caux_PRL_110}.

\begin{figure}
\begin{center}
\includegraphics[width=\columnwidth]{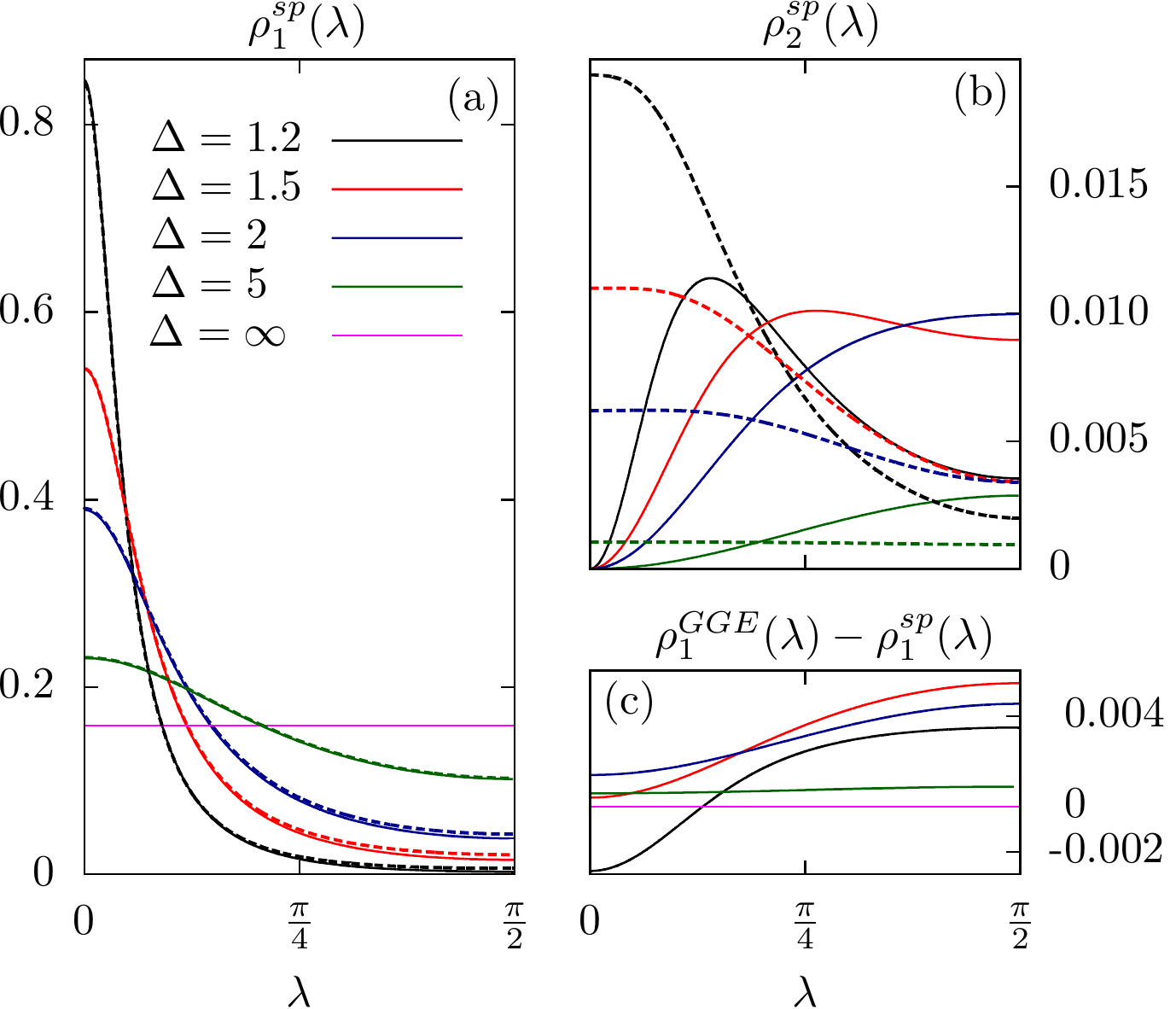}
\caption{\label{fig:rhos_gapped} (color online) (a),(b) Density functions 
$\rho_1$ and $\rho_2$ for the quench to different values of $\Delta>1$ 
of both the quench action saddle-point 
state (solid lines) and the GGE equilibrium state (dashed lines). (c) Difference between the GGE 
prediction for $\rho_{1}$ 
and the quench action saddle-point result. 
All distributions are symmetric functions of 
$\lambda \in [-\pi/2,\pi/2)$.}
\end{center}
\end{figure}

The implementation of the quench action approach to the N\'eel-to-XXZ quench proceeds as follows (see 
the Supplemental Material~\cite{SupMat} for details). One of the main ingredients is the leading-order behavior of the overlaps 
$\langle \blam | \Psi_{0} \rangle$ in the thermodynamic limit. It was proven in 
Refs.~\cite{2014_Brockmann_JPA,2014_Brockmann_npi} (starting from Ref.~\cite{1998_Tsuchiya,2012_Koslowski_JSTAT_P05021}) that 
only overlaps between $|\Psi_0\rangle$ and parity-invariant Bethe states are nonvanishing; practical determinant 
expressions were also derived. Taking $M$ to be even, rapidities of parity-invariant states come in pairs such that 
$\{\lambda_j\}_{j=1}^M= \{-\lambda_j\}_{j=1}^M$ and the overlap is now determined by $M/2$ rapidities 
$\tilde{\blam}=\{\lambda_j\}_{j=1}^{M/2}$. The overlap's leading term (in system size) reads \cite{2014_Brockmann_JPA}
\begin{equation} \label{eq:overlapsXXZleading}
\langle \tilde{\blam} |\Psi_0 \rangle  \thicksim\prod_{j=1}^{M/2} \frac{\sqrt{\tan(\lambda_j+i\eta/2) 
\tan(\lambda_j-i\eta/2)}}{2\sin(2\lambda_j)} \epp
\end{equation}
One can straightforwardly separate the contributions of different string lengths and derive an expression 
for the thermodynamic overlap coefficients $S[\brho]$. Before varying the quench action, per 
Eqs.~\eqref{eq:variation}, one needs to add a Lagrange multiplier fixing the filling of the saddle-point state 
to the N\'eel state's $\limth M/N = 1/2$. Variation leads to a set of generalized thermodynamic 
Bethe ansatz (GTBA) equations for the functions $\eta_n$ \cite{SupMat},
\begin{equation}\label{eq:TBA_XXZ_fact}
	\ln(\eta_n) = d_n + s \ast \big[\ln(1+\eta_{n-1})+\ln(1+\eta_{n+1})\big]\epc
\end{equation}
where $n\geq 1$, $\eta_0(\lam)= 0$ by convention, and 
\begin{align}\label{eq:TBA_XXZ_fact_driving}
d_n(\lam) = \sum_{k\in\mathbb{Z}} e^{-2ik\lam}\frac{\tanh(\eta k)}{k}\left[(-1)^n-(-1)^k\right]\epp
\end{align}
The solution to the GTBA Eqs.~\eqref{eq:TBA_XXZ_fact}, substituted into the Bethe Eqs.~\eqref{eq:BTGthlim_fact}, 
leads to a set of root densities $\brho^\text{sp}$ describing the steady state of the N\'eel-to-XXZ quench. 
They can be numerically computed by truncating the infinite sets of Eqs.~\eqref{eq:TBA_XXZ_fact} and 
\eqref{eq:BTGthlim_fact}. In Figs.~\ref{fig:rhos_gapped}(a) and \ref{fig:rhos_gapped}(b), we plot saddle-point 
distributions of $1$- and $2$-strings for different values of $\Delta$. 
A notable feature is the vanishing of the even-length string densities at $\lam=0$, which corresponds 
to the fact that the overlaps \eqref{eq:overlapsXXZleading} between the N\'eel state and parity-invariant 
Bethe states with a string of even length centered at zero identically vanish.
Furthermore, for large $\Delta$ values, the density of $1$-strings becomes increasingly dominant, approaching the ground state of the Ising model [$\rho_1(\lam)=1/(2\pi)$ and $\rho_n(\lam)=0$ for $n\geq 2$], in accordance with the expected result for the quenchless point $\Delta=\infty$.

\paragraph*{NLCE.}
Our NLCE follows on Ref.~\cite{rigol_14} and has been tailored to solve the specific quench
studied in this work \cite{rigolpre_14,SupMat}. NLCEs enable the calculation of the infinite-time average 
(also known as the diagonal ensemble result) of correlation functions after the quench in the thermodynamic 
limit \cite{rigol_14,rigolpre_14}. The idea is that any spin-spin correlation can be computed as a sum 
over the contributions from all connected clusters $c$ that can be embedded on the lattice,
$\langle\sigma^z_i\sigma^z_j\rangle_\text{NLCE}=\sum_{c}M(c)\times \mathcal{W}_{\sigma^z_i\sigma^z_j}(c)$,
where $M(c)$ is the number of embeddings of $c$ per site, and 
$\mathcal{W}_{\sigma^z_i\sigma^z_j}(c)$ is the weight of $\sigma^z_i\sigma^z_j$ in $c$. 
The latter is calculated using the inclusion-exclusion principle 
$\mathcal{W}_{\sigma^z_i\sigma^z_j}(c)=\langle\sigma^z_i\sigma^z_j\rangle^\text{DE}_c-
\sum_{s \subset c} \mathcal{W}_{\sigma^z_i\sigma^z_j}(s)$, where the last sum runs over all connected subclusters 
of $c$, and $\langle\sigma^z_i\sigma^z_j\rangle^\text{DE}_c={\textrm{Tr} [\sigma^z_i\sigma^z_j\hat{\rho}^\text{DE}_c]}/
{\textrm{Tr} [\hat{\rho}^\text{DE}_c]}$ is the expectation value of $\sigma^z_i\sigma^z_j$ calculated 
with the density matrix in the diagonal ensemble $\hat{\rho}^\text{DE}_c$ (in cluster $c$).
In order to accelerate the convergence of the NLCE, we use Wynn's and Brezinski's resummation 
algorithms \cite{SupMat,rigol_bryant_06,rigol_bryant_07}.

\paragraph*{GGE.}
The integrable structure of the XXZ spin chain provides, in the thermodynamic limit, an infinite set of 
local conserved charges ${Q}_{m}$, $m\in\mathbb{N}$, such that $Q_1 \propto P$,  $Q_2 \propto H$
\cite{1994_Grabowski_MPLA_9,1995_Grabowski_AP_243}. For integrable models, it is conjectured (and shown for
specific quenches) that the steady state after a quench can be described by a GGE. For the XXZ spin chain, 
the latter is given by a set of densities $\brho^\text{GGE}$ that maximizes the Yang-Yang entropy 
$S_{\text{YY}}[\boldsymbol{\rho}]$ under the constraint of fixed expectation values of the local conserved 
charges $Q_m$ \cite{2012_Mossel_JPA_45, 2012_Caux_PRL_109}. This translates into GTBA equations of the same 
form as Eqs.~\eqref{eq:TBA_XXZ_fact} but now with the driving function $d_1$ determined by the chemical 
potentials associated with the charges and the remaining $d_n(\lam)=0$ for $n\geq 2$. 
Together with Eqs.~\eqref{eq:BTGthlim_fact}, this uniquely determines $\brho^\text{GGE}$. In general, the values 
of the chemical potentials are inaccessible for the XXZ model, except for a truncated GGE when only 
a small number of conserved charges is taken into account \cite{2013_Pozsgay_Jstat_10}. 

However, it turns out that the expectation values of all local conserved charges $Q_m$ on the initial 
state are in one-to-one correspondence with the density $\rho_{1,h}$ of $1$-string holes, i.e.,
\begin{equation}\label{eq:pinning_rho1h}
\limth \left( \frac{\left\langle \Psi_0 \right| Q_{2m+2} \left| \Psi_0 \right\rangle}{N \sinh^{2m+1}(\eta)} \right)  
= \sum_{k\in\mathbb{Z}}  \frac{ \hat{\rho}_{1,h}(k) -  e^{-|k|\eta }}{2\cosh(k \eta)} (ik)^{2m} \epc
\end{equation}
with $m\geq 0$ and $\hat{\rho}_{1,h}$ the Fourier transform of $\rho_{1,h}$, see Ref.~\cite{SupMat}. In the case 
of the N\'eel-to-XXZ quench, the expectation values of the conserved charges on the initial state 
\cite{2013_Fagotti_JSTAT_P07012} fix $\rho_{1,h}$ unambiguously \cite{FollowUp},
\begin{equation}\label{eq:rho1h_exact}
\rho_{1,h}^{\text{N\'eel}}(\lambda) = \frac{\pi ^2 a_1^3(\lambda) \sin ^2(2\lambda)}
{\pi^2 a_1^2(\lambda) \sin^2(2\lambda)+\cosh^2(\eta)} \epc
\end{equation}
where $a_1$ was defined right after Eq.~\eqref{eq:total_energy}. 
This makes the input from the chemical potentials redundant. The densities 
$\brho^\text{GGE}$ for the GGE can be found by solving the GTBA Eqs.~\eqref{eq:TBA_XXZ_fact} 
for $n\geq2$ [$d_n(\lam)=0$] and the Bethe Eqs.~\eqref{eq:BTGthlim_fact} with the constraint 
$\rho^\text{GGE}_{1,h} = \rho_{1,h}^{\text{N\'eel}}$.
\begin{figure}
\begin{center}
\includegraphics[width=\columnwidth]{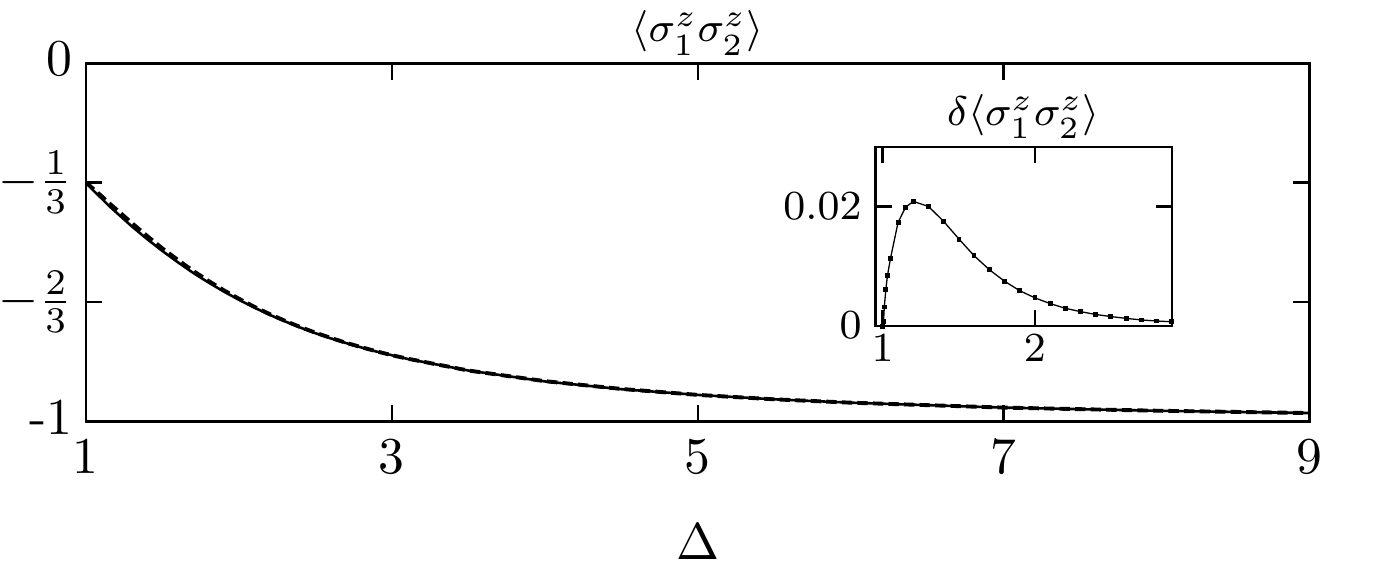}
\caption{\label{fig:corr_funcs} Correlator $\langle \sigma_1^z\sigma_2^z\rangle$ 
evaluated on the quench action steady state (solid lines) and on the GGE (dashed lines). 
The energy sum rule
$2\langle \sigma_1^x\sigma_2^x\rangle + \Delta\langle \sigma_1^z\sigma_2^z\rangle = -\Delta$
explains the exact value of $-1/3$ at the isotropic point $\Delta=1$.
Numerical errors are $10^{-5}$ or smaller. Both sets of data are in agreement with the 
finite-size computations of Ref.~\cite{2014_Fagotti_PRB_89}, within the numerical precision of the latter. Inset: Relative difference between the GGE prediction and the quench action saddle-point result, 
$\delta \langle \sigma_1^z\sigma_2^z\rangle = \left( \langle \sigma_1^z\sigma_2^z\rangle_\text{GGE} - \langle \sigma_1^z\sigma_2^z\rangle_\text{sp} \right)/\left| \langle \sigma_1^z\sigma_2^z\rangle_\text{sp} \right|$.}
\end{center}
\end{figure}

\paragraph*{Discussion of results.}
Numerical and analytical analysis show exact agreement between $\rho_{1,h}$ predicted by 
the quench action approach and $\rho_{1,h}^{\text{N\'eel}}$ in Eq.~\eqref{eq:rho1h_exact} \cite{FollowUp}. The expectation 
values of all local conserved charges $Q_n$ are, thus, reproduced exactly. We stress that this nontrivial 
agreement constitutes strong evidence for the correctness of the quench action prediction of the steady state. 

Furthermore, the distributions of the GGE can be compared with the steady-state distributions 
provided by the quench action approach, see Fig.~\ref{fig:rhos_gapped}. The 
densities $\rho_n$ for the GGE and the quench action are clearly different, the discrepancies becoming 
more pronounced as one reduces the anisotropy towards the gapless point $\Delta = 1$. 
We emphasize that all our results are obtained in the thermodynamic 
limit: these differences are not finite-size effects.

We verified the existence of these discrepancies by analytically solving the GTBA equations of the two ensembles 
in a large-$\Delta$ expansion. The differences between the distributions are of order $\Delta^{-2}$, e.g., 
for $1$- and $2$-strings (for other strings and higher orders, see Ref.~\cite{FollowUp})
\begin{subequations}\label{eq:diffdist}
\begin{align} 
\rho_{1}^\text{GGE}(\lam) - \rho_{1}^\text{sp}(\lam) &= \frac{1}{4\pi \Delta^2}+ O(\Delta^{-3}) \epc  \\
\rho_{2}^\text{GGE}(\lam) - \rho_{2}^\text{sp}(\lam) &=  \frac{1-3 \sin^2 (\lam)}{3\pi \Delta^2}  + O(\Delta^{-3}) \epp
\end{align}
\end{subequations}
\begin{figure}
\begin{center}
\includegraphics[width=\columnwidth]{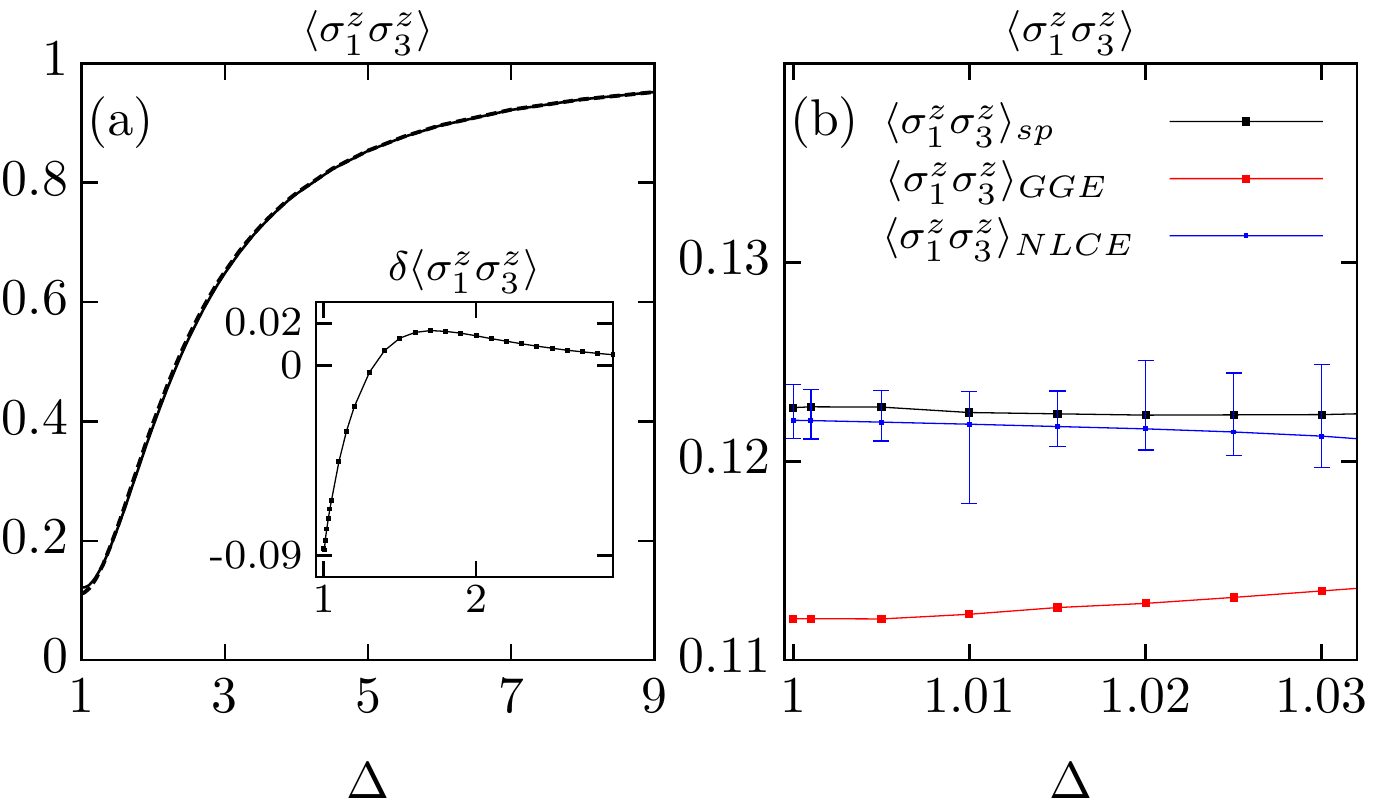}
\caption{\label{fig:nnn_correlator} (color online) (a) The same as Fig.~\ref{fig:corr_funcs} for $\langle \sigma_1^z\sigma_3^z\rangle$. (b) Comparison between the 
quench action, the GGE prediction, and the NLCE result close to the isotropic point. Error bars in the NLCE data display an interval of confidence that 
includes all resummation results (except for $\Delta\!=\!1.015$) \cite{SupMat}.}
\end{center}
\end{figure}

Given steady-state distributions, one can compute physical observables [Eq.~\eqref{eq:spev}]. 
Nonvanishing differences between distributions will generally be reflected in those expectation values, 
even in simple ones such as few-point spin-spin correlation functions. We have implemented an adapted 
version of the Hellmann-Feynman theorem to compute the expectation value 
$\langle \sigma_1^z \sigma_2^z \rangle$ from the distributions $\brho$ \cite{SupMat, 2014_Mestyan}. 
The nearest-neighbor two-point correlator is predicted differently by the quench action steady state and 
the GGE (see Fig.~\ref{fig:corr_funcs}). The NLCE results (not shown) are consistent with those predictions 
but cannot resolve their difference since it is too small ($\lesssim 2\%$, as
shown in the inset in Fig.~\ref{fig:corr_funcs}). It should be noted that the magnitude of differences 
between distributions in Eqs.~\eqref{eq:diffdist} does 
not directly translate into a similar difference for physical observables.
Expanding for large anisotropy, we obtain a 
discrepancy of order $\Delta^{-6}$,
\begin{equation}
\langle \sigma^z_1 \sigma^z_{2} \rangle_\text{GGE} - \langle \sigma^z_1 \sigma^z_{2} \rangle_\text{sp}  =  \frac{9}{16 \Delta^6} + O(\Delta^{-7}) \epp
\end{equation}

We also calculate the next-nearest-neighbor 
correlator $\langle \sigma_1^z \sigma_3^z \rangle$ by means of the method of Ref.~\cite{2014_Mestyan}, 
see Fig.~\ref{fig:nnn_correlator}. In the inset in Fig.~\ref{fig:nnn_correlator}(a) one can see 
that, as $\Delta\rightarrow1$, the differences between the predictions of the quench action approach 
and the GGE become of the order of $10\%$. Figure~\ref{fig:nnn_correlator}(b) provides a closer look of 
$\langle \sigma_1^z \sigma_3^z \rangle$ in that regime. There, we also report our NLCE results \cite{SupMat}.
The latter are consistent with the quench action predictions and inconsistent with the GGE results. 
Hence, our NLCE calculations support the correctness of the QA approach for describing observables 
after relaxation and the inability of the GGE constructed here to do so.

\paragraph*{Conclusions.}
We used the quench action method to obtain an exact description of the steady state following a quench 
from the ground state of the Ising model to an XXZ spin-1/2 chain with anisotropy $\Delta \geq 1$. 
We were also able to fully implement a GGE based on all known local conserved charges. 
Our main finding is that the quench 
action steady state is different from the GGE prediction. We have shown that even for local correlators, 
the methods produce different results.
An independent NLCE calculation supports the predictions of the quench action approach. 
A possible interpretation of our results is that GGE based on the local charges $Q_m$ 
is incomplete and that a larger set of conserved (quasi- or nonlocal) charges is needed 
\cite{2011_Prosen_PRL_106, 2014_Prosen_1406.2258, 2014_Pereira_1406.2306}. This makes it apparent 
that the study of quantum quenches provides a unique venue to further deepen our understanding of interacting integrable models.

It also remains an interesting open problem to extend our results to the gapless regime $-1<\Delta < 1$ and, going beyond steady-state issues, to reconstruct the postquench time-dependent relaxation itself, which is accessible via a quench action treatment and which would make correspondence to eventual experimental realizations more direct. We will return to these and further applications of the quench action method in future work.

\paragraph*{Acknowledgments.} We would like to thank P.~Ca\-la\-bre\-se, F.~Essler, M.~Fagotti, F.~G\"ohmann, V.~Gritsev, A.~Kl{\"u}mper, R.~Konik, M.~Kormos, B.~Pozsgay, and R.~Vlijm for useful discussions. We acknowledge support from the Foundation for Fundamental Research on Matter (FOM), the Netherlands Organisation for Scientific Research (NWO), and the US Office of Naval Research. For their support and hospitality, MB, MR and J-SC thank the Perimeter Institute, and BW, JDN and J-SC thank CUNY (where the main results of this work were first made public). This work forms part of the activities of the Delta Institute for Theoretical Physics (D-ITP).

\paragraph*{Note added.} 
After the first version of our preprint appeared, another preprint \cite{2014_Pozsgay} appeared that reports independent evidence that the quench action approach (and not the GGE based on known local conserved charges) correctly predicts the steady state for the N\'eel-to-XXZ quench. Our computations of the next-nearest-neighbor correlators were added after this, as well as the NLCE results. Two other related preprints \cite{2014_Andrei_1405.4224, 2014_Pozsgay_1406.4613} appeared in the intervening time.

%

\clearpage 
\setcounter{equation}{0}
\renewcommand{\theequation}{S\arabic{equation}}

\onecolumngrid

\begin{center}
{\Large Supplemental Material for EPAPS\\
Quenching the Anisotropic Heisenberg Chain: Exact Solution \\ and Generalized Gibbs Ensemble Predictions}
\end{center}

\section*{Thermodynamic limit of the overlaps}
In order to apply the quench action logic we need to compute the thermodynamic limit of the overlap coefficients and in particular their leading extensive part 
\begin{equation}
S[\boldsymbol{\rho}]= \limth S_{\boldsymbol{\lambda}} = - \limth \ln \frac{\langle \Psi_0 | \{\pm\lambda_j\}_{j=1}^{M/2} \rangle }{\|\{\pm\lambda_j\}_{j=1}^{M/2}\|} \epp
\end{equation}
The procedure is to consider the overlap coefficient for a generic finite size Bethe state $| \{ \lambda_j\}_{j=1}^M \rangle$ that in the thermodynamic limit, $N \to \infty$ with $ M/N=1/2$ fixed, flows to a set of distributions $| \{ \lambda_j\}_{j=1}^M \rangle \to | \boldsymbol{\rho} \rangle $. Equivalently, in the thermodynamic limit the eigenvalue of a smooth diagonal observable $\mathcal{A}$ can be recast into a sum of integrals weighted by the distributions $\boldsymbol{\rho}  = \{ \rho_n\}_{n=1}^\infty$:
\begin{equation}\label{eqn:therm_observable}
\mathcal{A} | \{ \lambda_j\}_{j=1}^M \rangle   =\Big[\sum_{j=1}^M A_j \Big]| \{ \lambda_j\}_{j=1}^M \rangle   \to \Big[N \sum_{n=1}^\infty \int_{- \pi/2}^{\pi/2} d\lambda \:\: \rho_n(\lambda) \tilde{A}_n(\lambda) \Big]  | \boldsymbol{\rho} \rangle \epp
\end{equation}
By assumption the extensive part of the overlap coefficient $S[\boldsymbol{\rho}]$ is smooth and does not depend on finite size differences within the set of Bethe states that scale to the same densities $\brho$. The number of Bethe states for each set of distributions $\boldsymbol{\rho}$ is given by the extensive Yang-Yang entropy \cite{SKorepinBOOK}: $e^{S_{YY}[\boldsymbol{\rho}]}$.  We are then free to select a representative finite size Bethe state out of all states that scale to the same $\brho$. We consider the state $| \{ \lambda_j\}_{j=1}^M \rangle $ consisting of $2n_s$ strings such that $2n_s = \sum_{n=1}^\infty M_n$, where $M_n$ is the number of $n$-strings in the state and we choose all $M_n$ to be even. Note that this is one possible choice for a representative state. Different choices regarding the eveness of the fillings $\{M_n \}_{n=1}^\infty$ lead to different expressions for the exact overlap formula \eqref{eqn:Overlap_Exact}, but are believed \cite{S2013_Caux_PRL_110,SFollowUp} to have the same extensive smooth part $S[\boldsymbol{\rho}]$.

For any finite size $N$, Bethe states are organized in deviated strings. We use the following notation to label the rapidities of such states:
\begin{equation}
\lambda_j \to \lam^{n,a}_\alpha =   \lambda^{n}_\alpha + \frac{i\eta}{2}(n+1 -2 a) + i\delta^{n, a}_\alpha \epc
\end{equation}
where $a= 1, \ldots n$ and $\alpha= 1, \ldots , M_n$. The string deviations $\delta^{n, a}_\alpha$ are vanishing in the thermodynamic limit. Though the string hypothesis is not systematically verified around the ground state of the zero-magnetized spin chain \cite{S1982_Woynarovich_JPA_15,S1983_Babelon_NPB_220_1}, it is effectively verified away from the ground state, for example at finite temperatures \cite{S1983_Tsvelik_AP_32}, and by extension in the circumstances discussed here.

The finite size overlap formula between this class of representative states and the N\'eel state is given in Ref.~\cite{S2014_Brockmann_JPA},
\begin{equation}\label{eqn:Overlap_Exact}
	\frac{\langle \Psi_0 | \{\pm\lambda_j\}_{j=1}^{M/2} \rangle }{\|\{\pm\lambda_j\}_{j=1}^{M/2}\|}=  \gamma \times  \sqrt{ \frac{\det_{M/2}(G_{jk}^{+})}{\det_{M/2}(G_{jk}^{-})}} \qquad \text{with} \quad \gamma = \prod_{j=1}^{M/2}\frac{\sqrt{\tan(\lambda_j+i\eta/2) \tan(\lambda_j-i\eta/2)}}{2\sin(2\lambda_j)} \epp
\end{equation}
The prefactor $\gamma$ has to leading order no explicit system size dependence from the string deviations $\delta \to 0$, but is exponentially vanishing when the particle number $M$ is sent to infinity due to the product over all rapidities. 
We focus on the ratio of the two determinants, where the matrices are given by
\begin{equation}
G^\pm_{(n,\alpha,a),(m,\beta,b)} = \delta_{(n,\alpha,a),(m,\beta,b)}\left(NK_{\eta/2}(\lambda^{n,a}_\alpha)-\sum_{(\ell,\gamma,c)}K_\eta^+(\lambda^{n,a}_\alpha, \lambda^{\ell,c}_\gamma)\right) + K_\eta^\pm(\lambda^{n,a}_\alpha ,\lambda^{m,b}_\beta) \epp
\end{equation}
Here, $K_\eta^\pm(\lambda,\mu)=K_\eta(\lambda-\mu) \pm K_\eta(\lambda+\mu)$ and $K_\eta(\lambda)=\frac{\sinh(2\eta)}{\sin(\lambda+i\eta)\sin(\lambda-i\eta)}$.
Divergences in system size as $1/\delta$ occur in each string block $(n=m,\alpha= \beta)$ when $b= a+1$ in the term $K_\eta(\lambda^{n,a}_\alpha - \lambda^{n,a+1}_\alpha) \sim \frac{i}{( \delta^{n, a+1}_\alpha -  \delta^{n, a}_\alpha )}$. On the other hand, for our representative state the terms $\pm K_\eta(\lambda+\mu)$ in $G^\pm$ are never divergent since all string centers are strictly positive. We conclude that the divergences in $1/\delta$ in $\det_{M/2}(G^+)$ cancel exactly the divergences in $\det_{M/2}(G^-)$, as they occur in exactly the same form. The same cancellation applies to divergences appearing in $K_\eta(\lambda- \mu)$ when two rapidities from different strings get close in the thermodynamic limit $\mu \to \lambda \pm i \eta + g(N)$ with $\limth g(N) = 0$. We are then able to take the thermodynamic limit $\limth$ for the overlap coefficients analogously to Ref.~\cite{S2014_DeNardis_PRA_89}. Being non-exponential in system size, the ratio of the two determinants can then be neglected in the contribution to the overlaps that is leading in system size. The thermodynamic overlap coefficients then read as
\begin{equation}
S[\brho] = \limth S_{\boldsymbol{\lambda}}  = -  \frac{N}{2} \sum_{n=1}^\infty \int_{0}^{\pi/2} \mathrm{d}\lam \, \rho_n(\lam) \ln W_n(\lam)  \epc
\end{equation}
where 
\begin{equation}
W_n (\lam) = \frac{1}{2^{n+1} \sin^{2}2\lam } \frac{ \cosh n \eta - \cos 2\lam }{\cosh n \eta + \cos 2\lam } \prod_{j=1}^{\frac{n-1}{2}} \left( \frac{ \cosh (2j-1) \eta - \cos 2\lam }{(\cosh (2j-1) \eta + \cos 2\lam) (\cosh 4\eta j - \cos 4\lam )}   \right)^{2} \epc
\end{equation}
if $n$ odd, and 
\begin{equation}
W_n (\lam) = \frac{ \tan^{2}\lam}{2^{n} } \frac{ \cosh n \eta - \cos 2\lam }{\cosh n \eta + \cos 2\lam }  \frac{1}{\prod_{j=1}^{\frac{n}{2}} \left(\cosh 2(2j-1)\eta - \cos 4\lam \right)^{2} } \prod_{j=1}^{\frac{n-2}{2}} \left( \frac{\cosh 2j\eta - \cos 2\lam }{\cosh 2j\eta + \cos 2\lam }\right)^{2} \epc
\end{equation}
if $n$ even.

\section*{GTBA equations for the N\'eel-to-XXZ quench}
In this section we focus on the derivation of the saddle point state, specified by the set of distribution $\brho^\text{sp}$ obtained by varying the quench action $S_{QA}[ \brho] = 2 S[\brho] -\tfrac{1}{2} S_{YY}[\brho]$ with respect to each root density $\rho_n(\lam).$ Variation of the Yang-Yang entropy is well-known \cite{SKorepinBOOK}. It should be noted that in front of the Yang-Yang entropy there is an extra factor $\tfrac12$. The reason is that only parity-invariant Bethe states contribute and therefore the number of microstates in the ensemble $\brho$ is the square root of the usual number of microstates. Furthermore, only states in the magnetization sector $M=N/2$ have non-zero overlap with the initial N\'eel state. In order to vary with respect to all $\rho_n(\lam)$ independently, we need to add a Lagrange multiplier term to the quench action:
\begin{equation}
-h\,N \left( \sum_{m=1}^{\infty} m \,  \int_{-\pi/2}^{\pi/2} \mathrm{d}\lam \, \rho_{m}(\lam) - \frac{1}{2}  \right) \epc
\end{equation}
where $h$ is the Lagrange multiplier. Variation with respect to $\rho_n(\lam)$ then leads to the saddle point conditions
\begin{equation} \label{eq:TBA_XXZ}
\ln \eta_{n}(\lam)  = - 2\,h\,n -  \ln W_{n}(\lam)  + \sum_{m=1}^{\infty}  a_{nm} \ast \ln \left( 1 + \eta_{m}^{-1} \right) (\lam) \epc
\end{equation}
where $n\geq1$, $\eta_n(\lam) = \rho_{n,h}(\lam)/\rho_n(\lam)$ and the convolution $\ast$ is defined by 
\begin{equation}\label{eq:convolution}
(f \ast g)(\lambda) = \int_{-\pi/2}^{\pi/2} d\mu \:  f(\lambda - \mu) g(\mu)
\end{equation}
The kernels are defined by $a_{nm}(\lam) =  (1-\delta_{nm})a_{|n-m|}(\lam) + 2 a_{|n-m| + 2}(\lam) + \ldots + 2 a_{n+m-2}(\lam) + a_{n+m}(\lam)$, where
\begin{equation}\label{eq:an}
a_{n}(\lam)  = \frac{1}{\pi} \frac{\sinh(n\eta)}{\cosh(n\eta) - \cos(2\lam)} \epp
\end{equation}
The functions $-2hn-\ln(W_n)$ are called driving terms. For each fixed value of $h$ the set \eqref{eq:TBA_XXZ} of Generalized Thermodynamic Bethe Ansatz (GTBA) equations has a solution in terms of the functions $\eta_n$. Substituting this solution into thermodynamic Bethe equations leads to the saddle point distributions $\brho^\text{sp}$. The parameter $h$ is fixed by the magnetization condition $M/N=1/2$ of the initial N\'eel state,
\begin{equation}
\sum_{m=1}^{\infty} m \,  \int_{-\pi/2}^{\pi/2} \mathrm{d}\lam \, \rho_{m}^\text{sp}(\lam) = \frac{1}{2} \epp
\end{equation}

As for the TBA equations at finite temperature \cite{STakahashiBOOK}, one can recast the GTBA Eqs.~\eqref{eq:TBA_XXZ} into a factorized form where there is no infinite sum over string  lengths. We will now derive this result. We use the Fourier transform conventions
\begin{align}\label{eq:FourierTransform}
	\hat{f}_k &= \ftk{f}{k} = \int_{-\pi/2}^{\pi/2}e^{2ik\lam}f(\lam)\,d\lam\epc\quad k\in\mathbb{Z}\epc \\
	f(\lam) &= \iftx{\hat{f}}{\lam} = \frac{1}{\pi}\sum_{k\in\mathbb{Z}} e^{-2ik\lam}\hat{f}_k\epp
\end{align}
The Fourier transforms of the kernels are $\hat{a}_{n,k} =e^{-|k|n\eta}$, and using the convolution theorem this implies $a_m \ast a_n = a_{m+n}$. A set of identities for the kernels can then be derived \cite{STakahashiBOOK}
\begin{subequations} \label{eq:kernelidentities}
\begin{equation}
	(a_0+a_2)\ast a_{nm} = a_1\ast (a_{n-1,m}+a_{n+1,m}) + (\delta_{n-1,m}+\delta_{n+1,m})\,a_1\epc \qquad n\geq 2,\, m \geq 1 \epc
\end{equation}
and
\begin{equation} \label{eq:2ndkernelidentity}
(a_0+a_2)\ast a_{1,m} = a_1 \ast a_{2,m} + a_1 \, \delta_{2,m}  \epc \qquad m\geq 1 \epc
\end{equation}
\end{subequations}
where we used the convention $a_0(\lam) = \delta(\lam)$. Convolving the GTBA Eqs.~\eqref{eq:TBA_XXZ} with $(a_0 + a_2)$, the infinite sum can be removed, and we find that
\begin{equation}\label{eq:gTBA_fact_prev}
	(a_0+a_2)\ast\ln(\eta_n) = (a_0+a_2)\ast g_n - a_1\ast(g_{n-1}+g_{n+1}) + a_1 \ast \big[\ln(1+\eta_{n-1})+\ln(1+\eta_{n+1})\big] \epp
\end{equation}
Here, the driving terms of the original GTBA equations are rewritten in a more convenient form with $g_n(\lam) = - \ln W_n(\lam) - 2n \ln 4$, where
\begin{subequations}\label{eq:gTBA}
\begin{align}\label{eq:gTBAa}
		g_n &= \sum_{l=0}^{n-1} \ln\left[\frac{s_{n-1-2l}c_{n-1-2l}s_{-n+1+2l}c_{-n+1+2l}}{t_{n-2l}t_{-n+2l}}\right]\epc \\
	t_n &= \frac{s_n}{c_n}\epc\quad s_n(\lam) = \sin\left(\lam+\frac{i\eta n}{2}\right)\epc\quad c_n(\lam) = \cos\left(\lam+\frac{i\eta n}{2}\right)\epp
\end{align}
\end{subequations}
Defining $g_0 (\lam) = 0$ and $\eta_0 (\lam) = 0$, Eq.~\eqref{eq:gTBA_fact_prev} holds for $n\geq1$. Let us rewrite the new driving terms $\tilde{d}_n =  (a_0+a_2)\ast g_n - a_1\ast(g_{n-1}+g_{n+1})$ of Eq.~\eqref{eq:gTBA_fact_prev}. We first rewrite $g_n$ such that only positive indices are present:
\begin{align}
	g_n = 2\delta_{n\,\text{mod}\,2,1}\ln\left[s_0^{(2)}\right] + 4 \sum_{l=1}^{\lfloor n/2 \rfloor}\ln\left[s_{n+1-2l}^{(2)}\right] + 2 \sum_{l=1}^{n-1}\ln\left[\frac{c_{l}^{(2)}}{s_{l}^{(2)}}\right] + \ln\left[\frac{c_{0}^{(2)}}{s_{0}^{(2)}}\right] 
+ \ln\left[\frac{c_{n}^{(2)}}{s_{n}^{(2)}}\right] \epc
\end{align}
where $s_l^{(2)} = s_ls_{-l}$, $c_l^{(2)} = c_lc_{-l}$ or, explicitly,
\begin{subequations}
\begin{align}
	s_l^{(2)}(\lam) &= \sin\left(\lam+\frac{i\eta l}{2}\right)\sin\left(\lam-\frac{i\eta l}{2}\right) = \sin^2\left(\lam\right)+\sinh^2\left(\frac{\eta l}{2}\right)\epc\\
	c_l^{(2)}(\lam) &= \cos\left(\lam+\frac{i\eta l}{2}\right)\cos\left(\lam-\frac{i\eta l}{2}\right) = \cos^2\left(\lam\right)+\sinh^2\left(\frac{\eta l}{2}\right)\epp
\end{align}
\end{subequations}
Now we use that for $\tilde{a}_\alpha(\lam) = \frac{1}{2\pi}\frac{\sinh(2\alpha)}{\sin^2(\lam)+\sinh^2(\alpha)}$ and $f_\beta(\lam) = \ln\left[\sin^2(\lam)+\sinh^2(\beta)\right]$ the following relation holds ($\alpha,\beta >0$):
\begin{equation}
	\tilde{a}_{\alpha}\ast f_\beta = f_{\alpha+\beta}-2\alpha\epp
\end{equation}
This implies the identities
\begin{equation}
a_m\ast \ln\left[\frac{c_l^{(2)}}{s_l^{(2)}}\right] = \ln\left[\frac{c_{l+m}^{(2)}}{s_{l+m}^{(2)}}\right] \epc \quad a_l\ast \ln\left[\frac{s_{0}^{(2)}}{s_{2}^{(2)}}\right] = \ln\left[\frac{s_{l}^{(2)}}{s_{l+2}^{(2)}}\right]  \quad \text{and}\quad a_l\ast \ln\left[\frac{c_{0}^{(2)}}{c_{2}^{(2)}}\right] = \ln\left[\frac{c_{l}^{(2)}}{c_{l+2}^{(2)}}\right] \epp
\end{equation}
From this we can calculate $\tilde{d}_{2n}$ and $\tilde{d}_{2n-1}$ for all $n\geq 1$ explicitly:
\begin{equation}
	\tilde{d}_{2n} =  \ln\left[\frac{c_0^{(2)}}{c_2^{(2)}}\right] - \ln\left[\frac{s_0^{(2)}}{s_2^{(2)}}\right] \epc\qquad
	\tilde{d}_{2n-1} =\ln\left[\frac{c_0^{(2)}}{c_2^{(2)}}\right] + \ln\left[\frac{s_0^{(2)}}{s_2^{(2)}}\right] \epp
\end{equation}
The GTBA equations can be written compactly as
\begin{subequations}
\begin{equation}\label{eq:TBA_XXZ_fact0}
	(a_0+a_2)\ast\ln(\eta_n) = \tilde{d}_n + a_1 \ast \big[\ln(1+\eta_{n-1})+\ln(1+\eta_{n+1})\big]\epc
\end{equation}
where $n\geq1$, the $\lam$-dependence is left implicit and by convention $\eta_0(\lam)= 0$ and $a_0(\lam)=\delta(\lam)$. The driving terms are given by
\begin{equation}
\tilde{d}_{n}(\lam) = \ln\left[\frac{\cos^2(\lam)}{\cos^2(\lam)+\sinh^2(\eta)}\right] - (-1)^n \ln\left[\frac{\sin^2(\lam)}{\sin^2(\lam)+\sinh^2(\eta)}\right]\epp
\end{equation}
\end{subequations}
Using the convolution theorem once again, one can invert the operation of $(a_0+a_2)\ast$ and bring it to the right hand side of Eq.~\eqref{eq:TBA_XXZ_fact0}. The Fourier transforms of the driving terms are
\begin{equation}
\hat{\tilde{d}}_{n,k} = 2\pi \frac{(1-e^{-2\eta|k|})}{|k|}\left(\frac{(-1)^n-(-1)^k}{2}\right) \epp
\end{equation}
Defining 
\begin{align}
\hat{d}_{n,k} &:= \frac{\hat{\tilde{d}}_{n,k} }{\hat{a}_{0,k}+\hat{a}_{2,k}} = 2\pi \frac{\tanh(\eta k)}{k}\left(\frac{(-1)^n-(-1)^k}{2}\right)\epc \nonumber \\
	\hat{s}_k & := \frac{\hat{a}_{1,k}}{\hat{a}_{0,k}+\hat{a}_{2,k}} = \frac{1}{2\cosh(k\eta)} \epc
\end{align}
the GTBA equations in Fourier space are
\begin{equation}
	\ft{\ln(\eta_n)}(k) = \hat{d}_{n,k} + \hat{s}_k\Big( \ft{\ln(1+\eta_{n-1})}(k) + \ft{\ln(1+\eta_{n+1})}(k) \Big) \epp
\end{equation}
In real space they are precisely the GTBA equations of the main text.

\section*{Nearest-neighbor spin-spin correlation}\label{sec:nn-correlator}
In this section we show how to compute the expectation value of the nearest-neighbor spin-spin correlator, $\langle\sigma_j^z \sigma_{j+1}^z\rangle = 1 + \frac{4}{N J} \left\langle \frac{\partial H}{\partial \Delta} \right\rangle$, in the thermodynamic limit on a generic, translationally invariant Bethe eigenstate specified by a set of smooth distributions $\boldsymbol{\rho}$. This is a direct consequence of the Hellmann-Feynman theorem \cite{S1939_Feynman}. Independently, an analogous implementation was recently presented in Ref.~\cite{S2014_Mestyan}.

A naive application of the Hellmann-Feynman theorem on the saddle point state leads to a wrong result
\begin{equation}
\langle \boldsymbol{\rho}^\text{sp} |\frac{1}{N}\frac{\partial H }{\partial \Delta}  |  \boldsymbol{\rho}^\text{sp}   \rangle \neq \frac{\partial }{\partial \Delta}  \langle \boldsymbol{\rho}^\text{sp} | \frac{1}{N} H |  \boldsymbol{\rho}^\text{sp}   \rangle = -\frac{J}{2}  \epp
\end{equation}
This is because the overlaps of two Bethe states with different values of $\Delta$ are exponentially small in system size, which makes the thermodynamic limit and the derivative with respect to $\Delta$ noncommuting,
\begin{equation}
\lim_{\delta \to 0} \limth \langle \blam(\Delta) | \blam(\Delta + \delta) \rangle  \neq \limth \lim_{\delta \to 0} \langle \blam(\Delta)   | \blam(\Delta + \delta) \rangle \epp
\end{equation}
In order to apply the Hellmann-Feynman theorem we need to take the derivative of the energy eigenvalue of a generic Bethe state at finite size $N$ and then take the thermodynamic limit. Under the string hypothesis the energy eigenvalue $\omega_\blam$ becomes a function only of the string centers $ \lambda_\alpha^n$ of the Bethe state, since the string deviations vanish exponentially in system size,
\begin{equation}\label{eq:energy}
\omega_\blam = - \pi J \sinh (\eta) \sum_{n , \alpha}  \: a_{n}(\lambda_{\alpha}^n) \epc
\end{equation}
where $a_n$ is defined in Eq.~\eqref{eq:an}. We can now apply the Hellmann-Feynman theorem to this finite size state by taking the derivative of the energy with respect to $\Delta$:
\begin{equation}\label{eq:corr1}
\langle \blam |  \sigma^z_1 \sigma^z_{2} | \blam \rangle = 1+ \frac{4}{NJ} \frac{d\omega_\blam}{d \Delta}  =1+ \frac{4}{NJ} \frac{1}{\sinh \eta}\frac{d\omega_\blam}{d \eta} = 1 - \frac{4\pi}{N} \sum_{n , \alpha} \left[ \frac{\cosh \eta}{\sinh \eta} a_{n}(\lambda_{\alpha}^n) + (\partial_{\eta} a_{n})(\lambda_{\alpha}^n) + \partial_{\lambda_{\alpha}^n} a_{n}(\lambda_{\alpha}^n) \frac{d \lambda_{\alpha}^n}{d \eta} \right] \epp
\end{equation}
The quantities $\frac{d \lambda_{\alpha}^n}{d \eta}$ are obtained by deriving the reduced Bethe equations for string centers \cite{S1931_Bethe_ZP_71,S1971_Takahashi_PTP_46}
\begin{equation}
\theta_{n}(\lambda_{\alpha}^n) - \frac{1}{N} \sum_{m ,\beta} \theta_{nm}(\lambda_{\alpha}^n - \lambda_{\beta}^m ) = \frac{2 \pi}{N} I_{\alpha}^n 
\end{equation}
(see \cite{STakahashiBOOK} for the definitions of $\theta_n$ and $\theta_{nm}$) with respect to the interaction parameter $\eta$,
\begin{equation}
N\, \tilde{a}_{n}(\lambda_{\alpha}^n) -  \sum_{m ,\beta} \tilde{a}_{nm}(\lambda_{\alpha}^n - \lambda_{\beta}^m )  + \sum_{m ,\beta} G_{(n, \alpha),(m, \beta)} \frac{d \lambda_{\beta}^m}{ d \eta} = 0 \epp
\end{equation}
Here we introduced the reduced Gaudin matrix for string centers \cite{S2005_Caux_JSTAT_P09003}
\begin{equation}
G_{(n, \alpha),(m, \beta)}  = \delta_{(n, \alpha),(m, \beta)} \Big( N\, a_{n}(\lambda_{\alpha}^n )  - \sum_{k, \gamma} a_{nk}(\lambda_{\alpha}^n - \lambda_{\gamma}^k ) \Big) +  a_{nm}(\lambda_{\alpha}^n - \lambda_{\beta}^m ) 
\end{equation}
and the functions $\tilde{a}_{n} $ and $\tilde{a}_{nm}$,
\begin{equation}
\tilde{a}_{n}  (\lambda) = \frac{1}{2 \pi }\partial_\eta \theta_n  (\lambda) =-  a_n(\lambda)   \frac{n \sin 2\lam}{2\sinh n \eta}\epc
\end{equation}
\begin{equation}
\tilde{a}_{nm}(\lambda) =  \frac{1}{2 \pi }\partial_\eta \theta_{nm} =  (1-\delta_{nm})\tilde{a}_{|n-m|}(\lambda) + 2 \tilde{a}_{|n-m| + 2}(\lambda) + \ldots + 2 \tilde{a}_{n+m-2}(\lambda) + \tilde{a}_{n+m}(\lambda)  \epp
\end{equation}
We define a set of the auxiliary functions $\boldsymbol{h}=\{ h_n\}_{n=1}^\infty$ such that
\begin{equation}\label{eq:def_hn}
{h_n(\lambda_\alpha^n)} = \sum_{m,\beta}  (G^{-1})_{(n, \alpha),(m, \beta)} \Big( N\, \tilde{a}_{m}(\lambda_{\beta}^m) -  \sum_{k  ,\gamma} \tilde{a}_{mk}(\lambda_{\beta}^m - \lambda_{\gamma}^k )  \Big) \epp
\end{equation}
In the thermodynamic limit, expression \eqref{eq:corr1} can then be recast as a functional of the distributions $\boldsymbol{\rho}$ and the auxiliary functions $\boldsymbol{h}$,
\begin{equation} \label{eq:rescorr}
\langle \brho | \sigma^z_1 \sigma^z_{2} | \brho \rangle = 1+ \limth \frac{4}{NJ}\frac{d\omega_\blam}{d \Delta}  =1  - 4\pi \sum_{n=1}^\infty \int_{-\pi/2}^{\pi/2} d\lambda \, \rho_n(\lambda) \left[ \frac{\cosh \eta}{\sinh \eta} a_{n}(\lambda) + \partial_{\eta} a_{n}(\lambda) - (\partial_{\lambda} a_{n}(\lambda)) {h_n(\lambda)}{ } \right] \epc
\end{equation}
where the auxiliary functions are determined by a set of linear integral equations
\begin{equation}
\rho_{n,t}(\lambda) h_n (\lambda) + \sum_{m=1}^\infty a_{n m} \ast ({\rho_m}{} h_m )\, (\lam)= \tilde{a}_n(\lambda) - \sum_{m=1}^\infty \tilde{a}_{nm}\ast \rho_m \,(\lam) \epp
\end{equation}
As in the case of GTBA Eqs.~\eqref{eq:TBA_XXZ}, they can be reduced to 
\begin{equation}
(a_0 + a_2) \ast (\rho_{n,t} h_n) = \Big[ (a_0 + a_2) \ast d_n  - a_1 \ast (d_{n-1} + d_{n+1})\Big] + a_1 \ast (h_{n-1} {\rho_{n-1,h}}{} + h_{n+1} {\rho_{n+1,h}}{}) \epc
\end{equation}
with $h_0(\lam) =d_0(\lam) = 0$ and the driving terms given by ($n \geq1$)
\begin{equation}
d_n(\lambda) = \tilde{a}_n(\lambda) - \sum_{m=1}^\infty \tilde{a}_{nm}\ast \rho_m\, (\lam) \epp
\end{equation}
Analogously to the local conserved charges, the correlator \eqref{eq:rescorr} can be expressed solely in terms of $\rho_{1,h}$ and the auxiliary function $h_1$ (see \cite{SFollowUp}),
\begin{align} 
\langle \brho | \sigma^z_1 \sigma^z_{2} | \brho \rangle  = 1+ 4\Big\{ & -\frac{\cosh \eta}{\sinh \eta} \sum_{k \in \mathbb{Z}}  \left(  \frac{ e^{- |k|\eta } - \hat{\rho}_{1,h}(k) }{2  \cosh k \eta} \right)  + \sum_{k\in\mathbb{Z}} |k | \left[\frac{ e^{- |k| \eta} }{2 \cosh k \eta} +\tanh(|k| \eta)  \left(   \frac{ e^{- |k|\eta } - \hat{\rho}_{1,h}(k) }{2  \cosh k \eta} \right) \right] \nonumber  \\
& - \pi \int_{-\pi/2}^{\pi/2} d\lambda  \: \rho_{1,h}(\lambda) \, h_1(\lambda)     \frac{\partial}{\partial \lambda}
s(\lambda) \Big\} \epc
\end{align}
where the function $s$ is given in the main text and $\hat{\rho}_{1,h}$ is the Fourier transform, defined in \eqref{eq:FourierTransform}, of $\rho_{1,h}$.

\section*{Numerical Linked-Cluster Expansions}
In this section we discuss the numerical linked-cluster expansion (NLCE) results. Such an expansion allows 
us to obtain the infinite-time average, or the diagonal ensemble (DE), result for the expectation 
value of spin-spin correlations. This approach was introduced in 
Ref.~\cite{Srigol_14} to study quenches with initial thermal states. 
It can be straightforwardly tailored to study quenches with initial ground states 
as explained in detail, for the particular quenches considered in this work, 
in Ref.~\cite{Srigolpre_14}. Here we report the results that are relevant to the 
discussion in the main text.

\begin{figure}[!b]
    \includegraphics[width=0.63\textwidth]{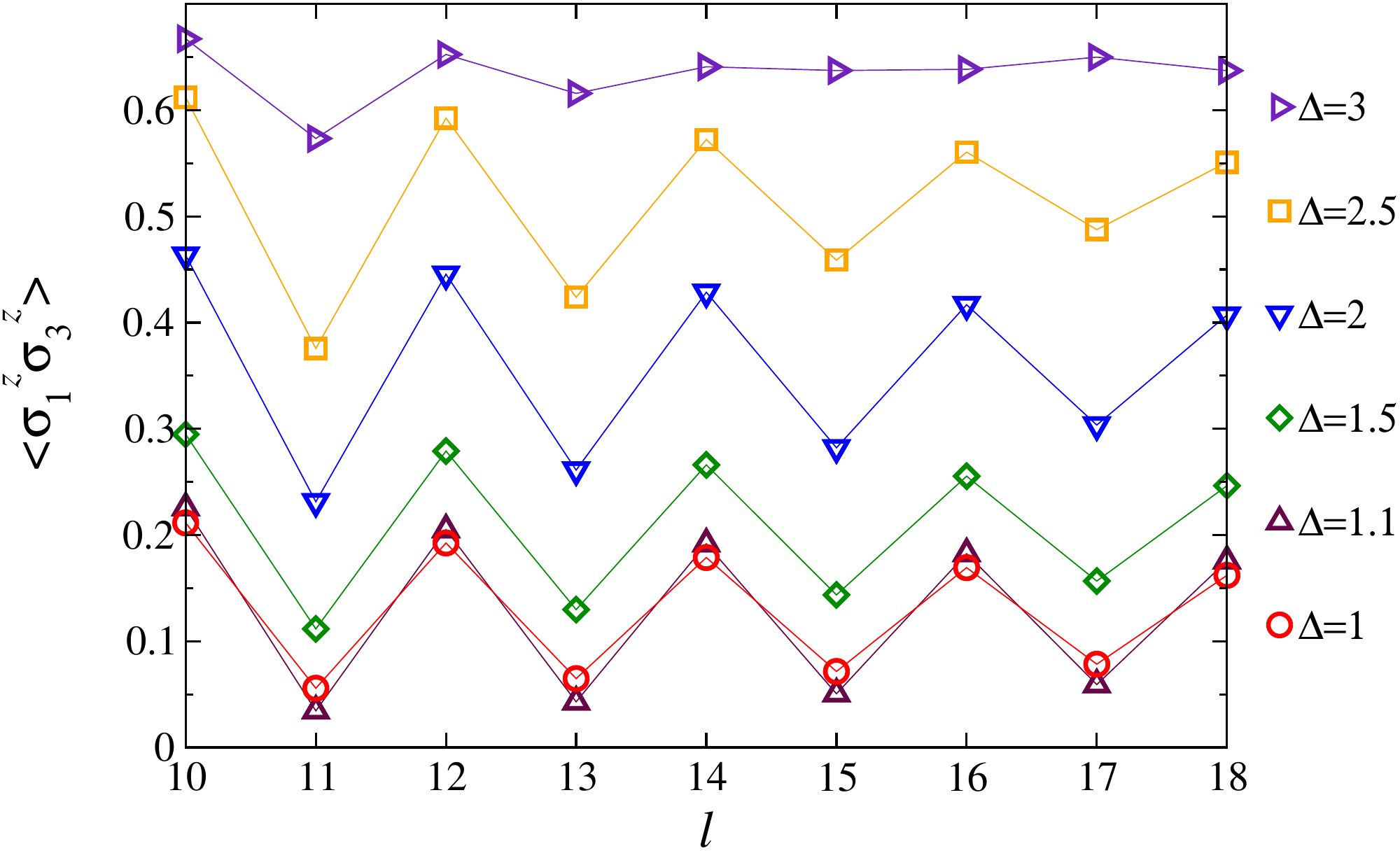}
\vspace{-0.1cm}
\caption{Next-nearest neighbor correlations $\langle\sigma^z_1\sigma^z_3\rangle^\text{DE}_l$
versus $l$ for quenches with $\Delta=1$, 1.1, 1.5, 2, 2.5, and 3. Results are reported 
for the last nine orders of the NLCE.}\label{fig:NLCELargeDbarevsQA}
\end{figure}

The NLCE calculations are done in clusters with up to 18 sites, as in 
Ref.~\cite{Srigol_14}. We denote as $\mathcal{O}^\text{DE}_l$ 
the result obtained for an observable $\mathcal{O}$ when adding the contribution 
of all clusters with up to $l$ sites. We should stress that while the 
convergence of the NLCE calculations improves for increasing values of $\Delta$ \cite{Srigolpre_14},  for large $\Delta$ they do not 
allow us to discriminate between the results of the quench action (QA) and the GGE 
calculations, which are very close to each other. 
They also do not allow us to discriminate between the QA and GGE results
as $\Delta\rightarrow1$ when the observable considered is 
$\sigma^z_1\sigma^z_2$. In that regime, the QA and GGE results 
are also too close to each other, becoming identical for $\Delta=1$. As shown 
in Fig.~3 of the main text, the largest relative differences between the 
QA and GGE predictions are seen for $\langle\sigma^z_1\sigma^z_3\rangle$
when $\Delta\rightarrow1$. This is the observable and regime in which we 
focus our NLCE effort.

\begin{figure}[!t]
    \includegraphics[width=0.63\textwidth]{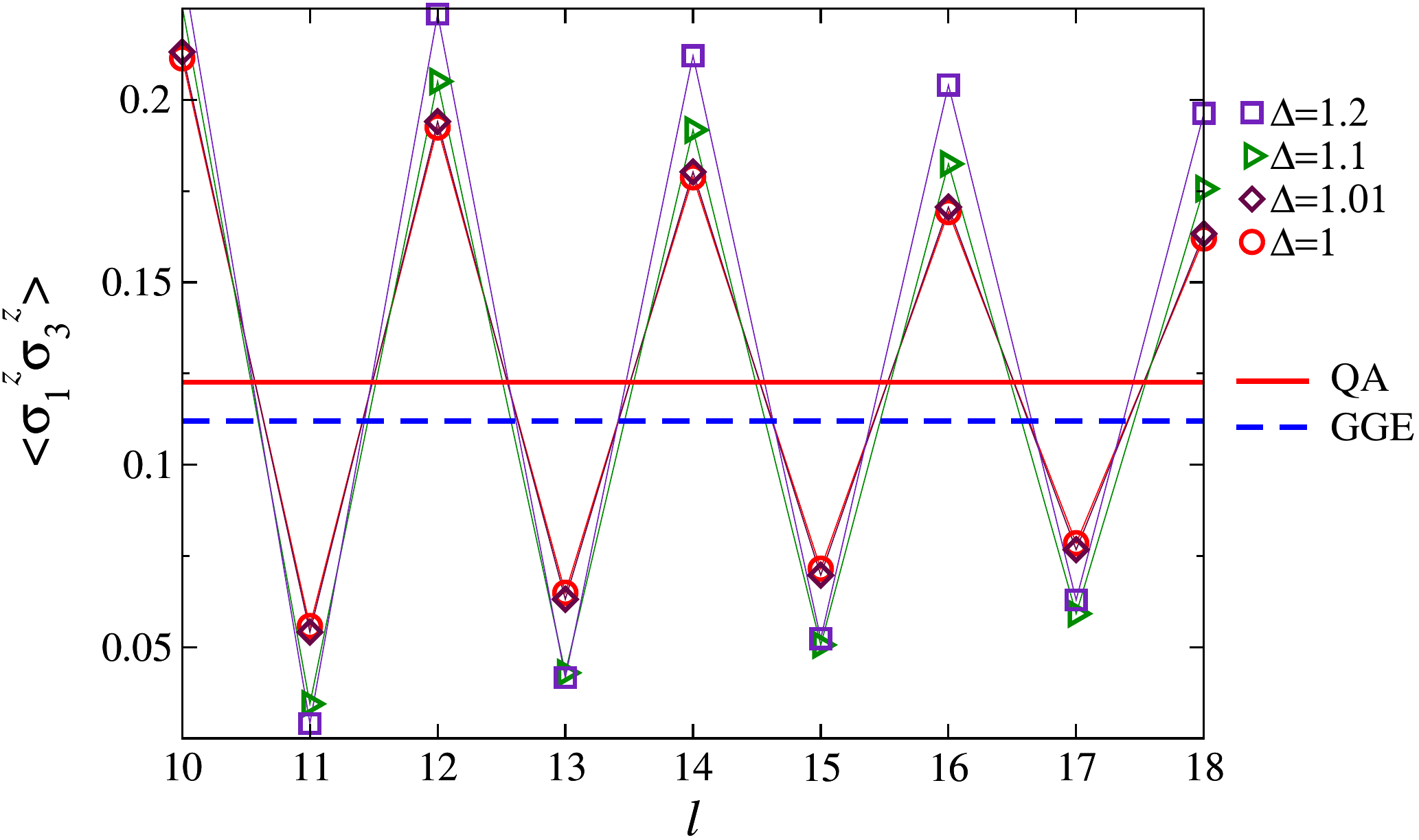}
\vspace{-0.1cm}
\caption{Next-nearest neighbor correlations $\langle\sigma^z_1\sigma^z_3\rangle^\text{DE}_l$
versus $l$ for quenches with $\Delta=1$, 1.01, 1.1, and 1.2. Results are reported 
for the last nine orders of the NLCE. The QA and GGE predictions for $\Delta=1$ are reported 
as continuous and dashed horizontal lines,
respectively.}\label{fig:NLCEbarevsQA}
\end{figure}

In Fig.~\ref{fig:NLCELargeDbarevsQA}, we show results for 
$\langle\sigma^z_1\sigma^z_3\rangle^\text{DE}_l$ versus $l$ for several values of 
$\Delta$ between 1 and 3. A few remarks are in order on those results. They can be
seen to be converging with increasing $l$, the amplitude of the oscillations decreases, 
but do not quite converge for the cluster sizes accessible to us. As $\Delta$ decreases 
from 3, one can see that the convergence of the NLCE initially worsens, the amplitude of 
the oscillations increases for any two contiguous values of $l$, and then improves very 
close to $\Delta=1$. Finally, it is 
apparent that the results for $\langle\sigma^z_1\sigma^z_3\rangle^\text{DE}$ decrease with 
decreasing $\Delta$, as expected from the exact calculations discussed in the main text.

We now turn our attention to the regime in which $\Delta$ is very close to 1. The fact 
that the convergence of our NLCE calculations improves in that regime is better seen in
Fig.~\ref{fig:NLCEbarevsQA}, where we report results for $\Delta=1$, 1.01, 1.1, and 1.2.
However, the results of our bare NLCE sums still do not allow us to discriminate between
the QA and GGE results for $\Delta=1$, which are depicted as continuous and dashed 
horizontal lines, respectively.
As discussed in NLCE studies of systems in thermal equilibrium
\cite{Srigol_bryant_06,Srigol_bryant_07}, whenever NLCE series do not converge to a desired
accuracy, one can use resummation techniques to accelerate the convergence of the 
series and obtain more accurate results. There are two resummation techniques that
were explained in detail in Ref.~\cite{Srigol_bryant_07}, which we have found to 
improve convergence in our problem. Those are Wynn's and Brezinski's algorithms. 
They can be applied multiple times (in what we call ``cycles'') to a series 
for an observable and are expected to improve convergence with each cycle.
However, the application of too many cycles can also lead to numerical 
instabilities \cite{Srigol_bryant_07}. 

\begin{figure}[!b]
    \includegraphics[width=0.54\textwidth]{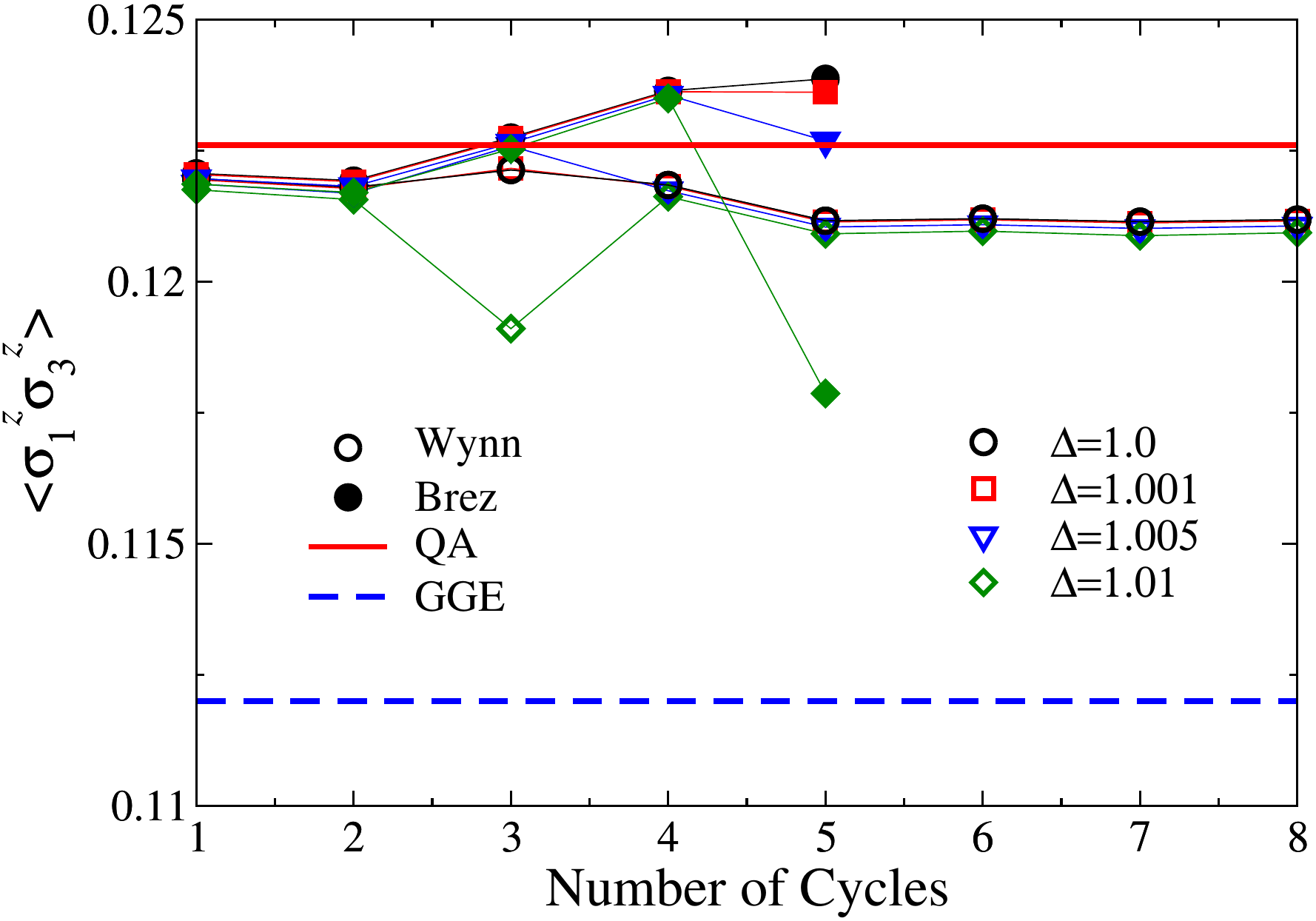}
\vspace{-0.1cm}
\caption{Next-nearest neighbor correlations $\langle\sigma^z_1\sigma^z_3\rangle^\text{DE}$
after each cycle of Wynn's (empty symbols) and Brezinski's (filled symbols) algorithms.
Results are reported for $\Delta=1$, 1.001, 1.005, and 1.01, and are compared to the QA 
(continuous line) and GGE (dashes line) results for $\Delta=1$.}
\label{fig:NLCEresumvsQA}
\end{figure}

In a nutshell, every cycle of those algorithms leads to a new series with fewer elements
\cite{Srigol_bryant_07}. The last element after each cycle is expected to approach the 
$l\rightarrow\infty$ result. The reduction in the number of elements after each cycle 
limits the number of times each algorithm can be applied to a series. In Wynn's algorithm, each
cycle reduces the number of elements by two, and in Brezinski's algorithm each cycle reduces the 
number of elements by three. Since our bare series for $\langle\sigma^z_1\sigma^z_3\rangle^\text{DE}$ 
has 18 elements (we consider clusters with up to 18 sites), the maximal number of cycles we can 
apply for Wynn's algorithm is 8 and for Brezinski's algorithm is 5.

In Fig.~\ref{fig:NLCEresumvsQA} we report the results (the last element) after each cycle of 
Wynn's and Brezinski's algorithms. As $\Delta$ approaches 1, we find that the resummations are 
stable and lead to very close results. This gives us confidence in the robustness of the  
resummations around the Heisenberg point. Figure~\ref{fig:NLCEresumvsQA} shows that,
as $\Delta\rightarrow1$ and as the number of cycles increases, Brezinski's algorithm 
appears to converge to a slightly larger 
value of $\langle\sigma^z_1\sigma^z_3\rangle^\text{DE}$ than the QA prediction, while Wynn's
resummations seem to converge to a result slightly below the QA prediction. Both are far 
from the GGE prediction. Hence, our NLCE results support the correctness of the QA predictions 
for the outcome of the relaxation dynamics in these systems. 

In the main text, we report the results of Brezinski's resummations after one cycle as representative 
of the outcome of the resummation techniques used. This because those results are almost identical to the ones 
obtained using Wynn's algorithm after one cycle, and are in between the results obtained after the 
maximal number of cycles that could be applied in each algorithm. In the main text, we also report an 
interval of confidence that contains all results obtained within Brezinski's and Wynn's resummations, 
except for $\Delta=0.015$ for which the last cycle of Brezinski's algorithm 
resulted in $\langle\sigma^z_1\sigma^z_3\rangle^\text{DE}=0.1346$, which is out of the interval 
of confidence reported. We should add that the fluctuations 
in the values of $\Delta$ after resummations increase as $\Delta$ increases for $\Delta<2$. Because
of this, and the fact that the QA and GGE results approach each other with increasing $\Delta$,
we cannot use our NLCE results to discriminate between the QA and GGE predictions as one moves 
further away from the Heisenberg point.

\end{document}